# A First Look at Identity Management Schemes on the Blockchain[*]

Paul Dunphy and Fabien A. P. Petitcolas,
Innovation Centre,
VASCO Data Security
{paul.dunphy,fabien.petitcolas}@vasco.com

**Abstract.** The emergence of distributed ledger technology (DLT) based upon a blockchain data structure, has given rise to new approaches to identity management that aim to upend dominant approaches to providing and consuming digital identities. These new approaches to identity management (IdM) propose to enhance decentralisation, transparency and user control in transactions that involve identity information; but, given the historical challenge to design IdM, can these new DLT-based schemes deliver on their lofty goals? We introduce the emerging landscape of DLT-based IdM, and evaluate three representative proposals – uPort, ShoCard and Sovrin – using the analytic lens of a seminal framework that characterises the nature of successful IdM schemes.

**Keywords:** Distributed Ledger Technology; Identity and Access Management.

## 1 Introduction

Twenty-four years have passed since Peter Steiner first showed the world that "on the Internet, nobody knows you're a dog", yet that famous drawing still stands to illustrate the challenge to identify individuals online. Today, we are very far from the public directory vision of the inventors of public-key cryptography in the seventies or the grand scheme of hierarchical certification envisaged in the eighties. Identity management (IdM) on the Internet still relies on what Cameron [1] called a decade ago a "patchwork of identity one-offs" comprising several types of IdM systems that are restricted to specific domains and do not interact much with each other. Centralised models of IdM currently face challenges due to the increasing regularity of data breaches that lead to reputation damage, identity fraud, but above all a loss of privacy for all concerned. These recurring events highlight a lack of control and ownership that end-users experience with their digital identities [2]–[4].

The investigation of alternative approaches to IdM is being led by initiatives that seek to expand the trustworthiness and reach of digital forms of identity. The United States' National Strategy for Trusted Identities in Cyberspace (NSTIC) aims to accelerate the development of novel technologies that can increase trust in online transactions [5]. Also, ID2020 seeks to leverage emerging digital technologies to expand the reach of legal identities (mirroring the United Nations' goals to "provide [by 2030] legal identity for all, including birth registration" [6]). The emergence of Bitcoin [7] has also inspired fresh thinking about digital identity due to its underpinning distributed ledger technology (DLT) not needing a central authority to validate transactions of its native cryptocurrency. Thus, a globally decentralised network is able to reach consensus on the current state of its book of transactions, the "ledger". The *distributed ledger* itself is *an append-only shared record of transactions that is maintained by entities on a peer-to-peer network*; while the often-cited "*blockchain*" is a cryptographic data structure often employed in DLT that is constructed through successive cryptographic hashing of blocks of transactions.

---

[*] To appear in IEEE Security and Privacy Magazine special issue on "Blockchain Security and Privacy" in 2018

Given that DLT is suited to assuring consensus, transparency, and integrity of the transactions that it contains, a number of benefits of applying DLT to IdM have already been proposed:

- **Decentralised** – Identity information is referenced in a ledger that no single central authority owns or control.
- **Tamper-resistant** – Historical activities in the DLT cannot be tampered with and transparency is given to all changes to that data.
- **Inclusiveness** – New ways to bootstrap user identity can be conceived that expand the reach of legal identities and reduce exclusion.
- **Cost saving** – Shared identity information can lead to cost savings for relying parties along with the potential to reduce volume of personal information that is replicated in databases.
- **User control** – Users cannot lose control of their digital identifiers if they lose access to the services of a particular identity provider/broker.

But, given these proposed benefits of incorporating DLT into future IdM schemes, is the path to new forms of DLT-based IdM really "inevitable" [2]?

## 2  Identity management on the blockchain?

Identity Management (IdM) encompasses the *processes and policies involved in managing the lifecycle of attributes in identities for a particular domain* [8]. Most IdM schemes today are centralised where a single entity controls the system. However, the generated identities themselves can be federated beyond a single organisation, as when governments issue national identity cards. In federated identity systems, users can use identity information established in one security domain to access another. Single sign-on schemes, such as Facebook Connect, can work this way. User-centric identity management places administration and control of identity information directly into the hands of individuals. Examples include password managers (e.g., 1Password, Less-Pass) that securely keep track of different website credentials.

Despite the different approaches, one function that is fundamental to IdM is securely binding together an *identifier*: a value that unambiguously distinguishes one user from another in a domain; and *attributes* (sometimes called certifications or claims): entitlements or properties of a user such as name, age, credit rating etc. The first steps taken to tailor the use of DLT for establishing a secure and decentralised identifier-attribute mapping were taken in the design of *Namecoin*: the longest surviving software fork of

Bitcoin. Namecoin provides a human-readable, decentralised and secure namespace for the ".bit" web domain. This achievement contradicted conventional wisdom that a naming system exhibiting all three characteristics could not be designed [9]. Blockstack [4] has extended Namecoin's scheme, to create a decentralised public key infrastructure (PKI): it registers bindings between a public key and a human readable identifier.

Recently, several decentralised identity schemes have emerged that extend beyond naming and aim to provide a more complete suite of IdM functions. However, until now, there has been no evaluation of these proposals. We were interested in whether DLT-based IdMs have potential to go beyond previous approaches, or would simply create new "identity one-offs".

### 2.1 Approach

We started our inquiry by searching for blueprints of DLT-based IdM proposals that were technically scrutable (e.g., white papers, open source software, etc.). We excluded schemes that only provided naming and found that all fell into one of two categories:

1 – *Self-sovereign Identity*: an identity that is owned and controlled by its owner without the need to rely on any external administrative authority and without the possibility that this identity can be taken away. It can be enabled by a decentralised identity eco-system that facilitates the recording and exchange of identity attributes, and the propagation of trust among participating entities. Examples include Sovrin, uPort and OneName;

2 – *Decentralised Trusted Identity*: an identity that is provided by a centralised service that performs identity proofing of users based upon existing trusted credentials (e.g. passport), and records identity attestations on a DLT for later validation by $3^{rd}$ parties. Examples include ShoCard, BitID, ID.me and IDchainZ.

In this paper, we focus on three particular DLT-based IdM schemes: uPort, ShoCard, and Sovrin. We chose these three schemes in particular because individually they serve as key exemplars of the prevalent design decisions and challenges found in their respective genres, and together serve a similar purpose for the broader landscape of DLT-based IdM. In addition, they have provided the most technical detail of their scheme designs and are either underpinned by sizeable online communities or have notable venture capital funding.

There is no definitive criterion to evaluate IdM schemes, so in order to generate early insights about individual schemes we leveraged an evaluation framework known as the "laws of identity" [1] which serve to pinpoint the successes and failures of digital identity systems. It is a widely-known framework, and represents a full spectrum of IdM concerns, encompassing security, privacy and user experience. Furthermore, the laws provide an inherent flexibility, which is ideal for application to the heterogeneous and early-stage DLT-based IdM schemes we evaluated. The laws themselves are as follows:

1 – **User control and consent** – Information that identifies the user should only be revealed with that user's consent.

2 – **Minimal disclosure for a constrained use** – Identity information should only be collected on a "need-to-know" basis and kept on a "need-to-retain" basis.

3 – **Justifiable parties** – Identity information should only be shared with parties that have a legitimate right to access identity information in a transaction.
4 – **Directed identity** – Support should be provided for sharing identity information publicly or in a more discreet way.
5 – **Design for a pluralism of operators and technology** – A solution must enable the inter-working of different identity schemes and credentials.
6 – **Human integration** – The user experience must be consistent with user needs and expectations so that users are able to understand the implications of their interactions with the system.
7 – **Consistent experience across contexts** – Users must be able to expect a consistent experience across different security contexts and technology platforms.

In the text that follows, where we refer to a specific law, we use bracket notation to reference the law number (e.g., (1), (5)).

## 3 uPort

*uPort* [3] is an open source decentralised identity framework that aims to provide "decentralised identity for all". Its use case is IdM for next generation decentralised applications (DApps) on the Ethereum DLT and for traditional centralised applications such as email and banking.

### 3.1 Design

A uPort identity is underpinned by the interactions between Ethereum *smart contracts*: bespoke code that can regulate the movement of data and *ether* (the native cryptocurrency) on Ethereum. Smart contracts are uniquely addressed by 160-bit hexadecimal identifiers, and, when invoked are executed by the Ethereum Virtual Machine (EVM) installed on every Ethereum node. Two smart contract templates designed by uPort comprise each uPort identity: *controller* and *proxy*. To create a new identity, a user's uPort mobile application creates a new asymmetric key pair and sends a transaction to Ethereum that creates an instantiation of a controller that contains a reference to the newly created public key. Then, a new proxy is created that contains a reference to the address of the just-created controller contract; only the controller contract can invoke functions of the proxy; a constraint that is specified in the controller and enforced by the EVM. The address of the proxy comprises the unique *uPort identifier* (uPortID) of a user. A user is free to create multiple uPortIDs that are unlinkable. Figure 1 provides an overview of an interaction between a uPortID and the smart contract of a decentralised application on Ethereum.

The private key that controls a uPortID is stored only on the user's mobile device. Therefore, an important aspect of uPort relates to its social recovery protocol for the event of loss or theft of the user's mobile device. For that, users must nominate the uPortIDs of trustees who can vote to replace the public key referenced in the controller with one proposed by the user in need; once a quorum is reached between those trustees on the new public key, the controller replaces the lost public key with the newly proposed public key. This process enables the user to maintain a persistent uPortID even after the loss of cryptographic keys.

A final aspect of the uPort scheme is its support for securely mapping identity attributes to a particular uPortID. The uPort *registry* is a smart contract that stores the global mapping of uPortIDs to identity attributes. Any entity can query the registry, however, only the owner of a specific uPortID can modify its respective attributes. Due to the inefficiency of storing large volumes of data in a smart contract, only the hash of the JSON attribute structure is stored in the registry. The data itself is stored on IPFS: a distributed file system where a file can be retrieved by its cryptographic hash.

### 3.2 Analysis

uPort has no central server and does not authenticate the owner of a uPortID; this passes the risk of unauthorised access to the local authentication methods on the user's mobile device. While the social recovery protocol provides one method to recover ownership of a lost or compromised uPortID, the trustees themselves could be one vector of attack since their own uPortIDs are openly linked to the user's uPortID; this transparency provides opportunities for collusion against a specific uPort user. If an attacker can compromise a uPort application and replace trustees unnoticed via the controller, the uPortID is compromised permanently. So while uPort does place more control over uPortIDs in the hands of its users – a plus for (1) – a layer of added complexity and responsibility is inevitably handed to users.

uPort does not require personal data disclosures to bootstrap an uPortID for a constrained use and also respects privacy in terms of the lack of inherent linkability between uPortIDs (2). However, the registry (if used) represents a point of centralisation that can be probed for information about identifiers and identity data. So while specific attributes within the attribute data structure can be individually encrypted, the overall JSON data structure is still visible which could leak meta-data about specific attributes or relationships with identity providers/relying parties. Thus there is a chance that over-reliance upon the registry can compromise privacy (3).

A commerce application can widely advertise its uPortID, but uPort provides no public directory to look up uPortIDs from arbitrary search criteria. Discreet disclosure of a uPortID is possible if a user creates new uPortIDs for each new relying party that they encounter and avoids the use of the registry (4). Although, since a uPortID equates to a smart contract, an honest but curious Ethereum node could discover even non-disclosed uPortIDs through analysis of the smart contract code stored at a given address to determine if it is a uPort template. More work is needed to discover whether non-disclosed uPortIDs are private in practice.

uPort does not perform any identity proofing but instead provides a framework for users to gather attributes from an eco-system of trust providers; uPort simply specifies the format of attributes that are stored in its registry. But as a consequence of the uPortID owner alone having write-access to their own respective part of the registry, a user can selectively discard negative attributes that they are given e.g. a poor credit score etc. (5).

The mobile application of uPort provides a consistent user experience across all usage contexts (7) due to the scanning of a QR code to initiate interactions with a relying party. However, the in-app education is not present relating to the implications of putting representations of personally identifiable information on a DLT designed to prioritize immutability and transparency of data (6). The area of user education will become pressing in this context as legislation such as European General Data Protection Regulations (GDPR) come into force in Europe.

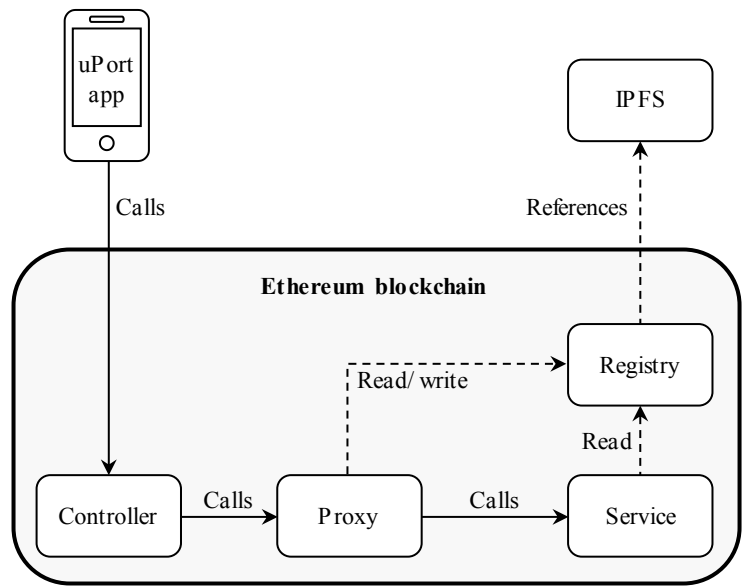

**Figure 1: An overview of key elements of uPort: a user identity is comprised of a mobile application and two smart contracts:** *controller* **and** *proxy*. **The registry is a smart contract that provides a decentralised mapping of uPort identifiers to identity attributes. The registry can be globally read, and can reference data stored in an off-DLT data store such as IPFS.**

## 4  Sovrin

Sovrin [2] is an open-source decentralised identity network built on permissioned DLT. Sovrin is public, but only trusted institutions, called *stewards* – which could be banks, universities, governments, etc. – can run nodes that take part in consensus protocols: thus the ledger is *permissioned*. The non-profit Sovrin Foundation ensures the proper governance of the *stewards* and their respect of a legal agreement called the *Sovrin Trust Framework*. Sovrin provides the code-base to the Hyperledger Indy project.

### 4.1 Design

Sovrin enables a user to generate as many identifiers as needed to keep contextual separation of identities for privacy purposes; each identifier is unlinkable and controlled by a different asymmetric key pair. Sovrin identifiers themselves are managed by the user or an appointed *guardian* service, and follow the Decentralised Identifier (DID) specification currently seeking IETF standardisation. A DID is a data-structure containing the user identifier, cryptographic public key and other meta-data necessary to transact with that identifier.

The Sovrin architecture can be summarised by the components as shown in Figure 2. The key element is the Sovrin ledger. This contains transactions associated with a specific identifier and is, written, distributed and replicated among the *steward* nodes, which run an enhanced version of the redundant Byzantine fault tolerant protocol of Aublin et al. [10], called Plenum, for consensus.

There are two important consequences to the choice of permissioned ledger in Sovrin's design. First, no expensive proof-of-work computation is required to reach consensus on the state of the ledger, reducing significantly the energy cost of running a node and improving dramatically transactions throughput. Second, trust on Sovrin relies on both people and code. Trust starts from the common root-of-trust formed by the globally distributed ledger, but as new organisations and users join the network, they can become *trust anchors* (i.e. allowed to add more users and organisations); a "web of trust" is expected to evolve to support this decentralised network growth.

Users interact with Sovrin through a mobile application and control software *agents* acting on their behalf to facilitate interactions with other *agents* on the network. *Agents* are network endpoints that are always addressable and accessible. Users could run *agents* on their own servers, but more likely, they will ask specialised intermediaries: *agencies*, to do that for them, alike e-mail systems. *Agents* also provide a backup service and encrypted storage of attribute credentials.

The mobile application also helps users manage cryptographic keys, which are stored on the users' mobile device. As in uPort, Sovrin offers a mechanism for key recovery that relies on the user selecting a set of trustees. When requested to do so by the user, a specified quorum of trustees must sign a new identity record transaction that *stewards* must verify.

### 4.2 Analysis

Sovrin aims to equip users to fully control all aspects of their identity. Each user can select from the attribute credentials that they hold about themselves, which they wish to share with a relying party (1). This is made possible through the use of anonymous credentials. Although users can choose to store those attributes on the ledger, in general, they will prefer to use the storage capabilities of their mobile phone or their *agent* to transmit attributes to other parties through secure communication channels, and use the ledger to identify the correct network endpoint to use. The use of attribute-based credentials allows users to only reveal credentials that they choose (2). Verifying the party with whom data is shared, remains a challenge, which is partly addressed through the web-of-trust, the governance of the Sovrin Foundation and the reputation of the *stewards*.

Although there are no trusted third parties in the PKI-sense on Sovrin, users must rely on *agencies* that will act on their behalf on the Sovrin network and on the *stewards* maintaining the ledger. Depending on the choice of agent and its implementation, a lot of information could potentially be in the hands of the agency. However, as agencies are acting on behalf of the user, they have a "necessary and justifiable place" in the identity relationship (3).

Sovrin supports both omnidirectional and unidirectional identifiers (4): public organisations can decide to publish their full identity on the network, while users may choose to publish only identifiers and to use different identifiers and cryptographic key-pairs with each party they interact with, avoiding emitting "correlation handles".

Today, Sovrin depends on a very small number of operators sharing the same implementation. As the systems gets traction, new agencies, and new *stewards*, will join. The Sovrin Foundation expects in particular to build a market of agencies that will compete on the features they offer, for instance interfaces with other (existing) identity systems (5).

Finally, one important issue not yet addressed by the Sovrin developers, is the user experience. The history of security offers several examples of smart cryptographic systems, which have never been deployed widely because users found it too cumbersome or difficult to understand – email encryption using PGP is a seminal example. So, human integration remains an open question for Sovrin. Considering that Sovrin is still in the early development phase, evaluating it against laws (6) to (7) is tricky, but it is illustrative that much work has considered the design of the scheme architecture itself, but hardly any has considered the user experience.

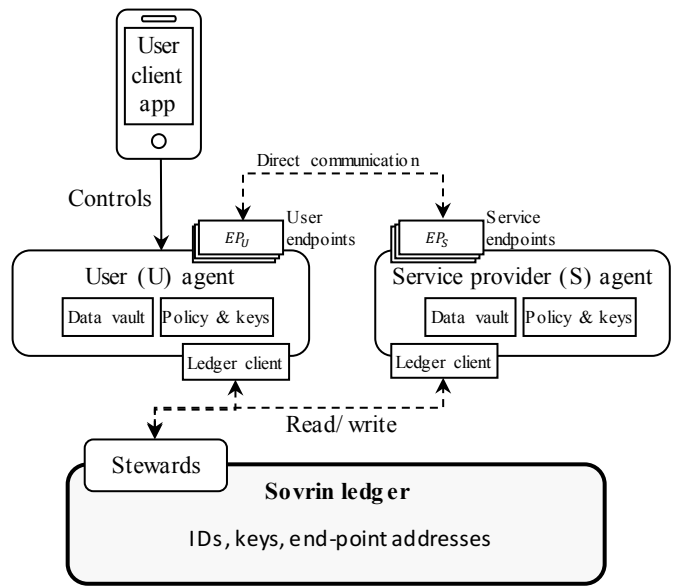

Figure 2: An overview of key elements of Sovrin. At the base of Sovrin is a permissioned ledger. Only *stewards* that legally abide by the *Sovrin Trust Framework* can write to the ledger. Users and organisations rely on *agents* that are addressable network points. Identifiers, keys and endpoint addresses are stored on the ledger.

## 5 ShoCard

ShoCard [11] provides a trusted identity that leverages DLT to bind a user identifier, an existing trusted credential (e.g., passport, driver's license), and additional identity attributes, together via cryptographic hashes stored in Bitcoin transactions. ShoCard's primary use cases are verification of identity in face-to-face as well as online interactions.

### 5.1 Design

ShoCard uses Bitcoin as a timestamping service for signed cryptographic hashes of the user's identity information, which are mined into the Bitcoin blockchain. ShoCard incorporates a central server as an essential part of its scheme; this server intermediates the exchange of encrypted identity information between a user and a relying party. The scheme relies on three phases: *bootstrapping*, *certification*, and *validation*. Figure 3 schematises those phases.

*Bootstrapping* occurs at the creation of a new ShoCard. The ShoCard mobile application generates a new asymmetric key pair for the user and scans their identity credentials using the device's camera. The scan and the corresponding data are encrypted and stored on the mobile device; the signed hash of this data is also embedded into a

Bitcoin transaction for later data validation purposes. The resulting Bitcoin transaction number constitutes the user's ShoCardID and is retained in the mobile application as a pointer to the ShoCard *seal*

Once a ShoCard is bootstrapped, the user can interact with identity providers to gather additional attributes in a process called *certification*. In order to associate certificates to a ShoCardID, an identity provider must first verify that the user knows both the data hashed to create it, and the cryptographic keys that signed the seal. In a face-to-face context, this can be achieved by the user providing the original identity data forming the seal from their mobile device, a digitally signed challenge and presenting the original trusted credential. The certificate takes the form of a signed hash of new attributes (and its associated ShoCardID) in a Bitcoin transaction created by the provider. The provider must share the Bitcoin transaction number along with a signed plaintext of the new attributes directly with the user. Since the user will later need to provide the attributes to relying parties and may not want to lose them if the mobile device is lost, a ShoCard server offers storage for encrypted version of certifications (known as an *envelope*). ShoCard never learns the encryption key, which enables the user to share certifications only with selected parties.

The *validation* phase occurs when a relying party must verify a certification to determine whether a user is entitled to access a service (e.g. has checked in to a flight). To validate the envelope the user must first provide the relying party with the envelope reference and its encryption key. After retrieving the envelope from the ShoCard servers, the relying party performs a number of checks: i) that the envelope signature was produced with the same private key that signed the seal; ii) that the certification signature was created by a trusted entity and the plain-text certification corresponds to the one hashed and signed in the certification; iii) finally, that the identity data presented by the user in the pending transaction match those signed and hashed in the seal.

### 5.2 Analysis

The ShoCard central server functions as an intermediary to manage the distribution of encrypted certifications between ShoCard users and relying parties. In this way, ShoCard bears less risk of data breach than if it stored and distributed plaintext identity data. Secure storage of identity information and appropriate sharing with relying parties is controlled by the end-user (1). However, ShoCard's intermediary role does create uncertainty about the longitudinal existence of a ShoCardID; if the company ceased to exist, users of ShoCard would be unable to use the system with the certifications they had acquired. This makes ShoCard more centralised in practice than its open reliance on DLT might suggest.

Each ShoCard identity is bootstrapped with an existing trusted credential, such as a passport or driving license. Such an approach may require users to embed more personal information in their ShoCard seal than they had originally intended. This may make ShoCard less attractive for low value online accounts (2).

Since the user is in control of initiating sharing activities, and since ShoCard only stores encrypted data, there can be some confidence that only justifiable parties are involved in the identity data sharing transaction. However, the ShoCard server may be able to associate a particular ShoCardID with a particular relying party, since envelopes must be retrieved from ShoCard server by the relying party (3).

ShoCard only supports unidirectional identifiers and does not have the concept of a public registry of ShoCardIDs. Although omnidirectional identifiers may be needed in future to realise its vision of an eco-system of reusable certifications (4).

ShoCard does support a multitude of different identity providers through its certification functionality, but those providers must create bespoke integration with ShoCard's own web services in addition to Bitcoin, which could be a barrier to uptake. The decision to perform such integration could be driven by the trustworthiness of ShoCard's identity proofing of its users (5).

The scanning of identity documents and QR codes is a dominant interaction paradigm in the ShoCard user experience: it is simple and consistent (7). However, it is unclear what the user motivations would be to adopt this new type of digital identity, and how users would be educated about the implications of referencing identity data on a blockchain (6). Users are also not supported with cryptographic key management.

One final point concerns the overall deployability of ShoCard. Bitcoin transactions take on average 10 minutes to be mined into the blockchain, and furthermore it is recommended to wait for six additional blocks to be mined before assuming the settlement of a transaction. This could bring the waiting time for settlement to one-hour on average. If a context that depends upon real-time settlement of certifications, this speed could create challenges for the user experience and adoption by vendors.

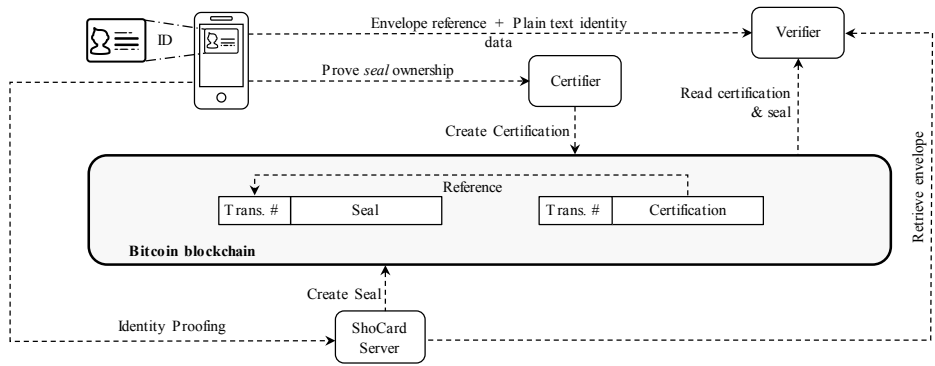

**Figure 3: An overview of key elements of ShoCard. ShoCard uses Bitcoin to record a commitment to personal data that was verified during identity proofing, and for storage of hashes of certifications that build upon the user's Seal created by relying parties. The ShoCard server plays an active role as an intermediary between users and relying parties.**

## 6 Discussion

Table 1 summarizes each scheme that we evaluated with respect to each law of identity. An unshaded table cell indicates that we found evidence that suggests a scheme complied with a specific law, and a shaded cell indicates that we didn't find evidence that suggested compliance with a specific law. We additionally include a summary of Facebook Connect to provide contrast.

### 6.1 Decentralisation that relies upon centralisation and intermediaries

DLT is often seen as a remedy for system architectures dominated by central authorities and intermediaries. But while each DLT-based IdM scheme we looked at leverages techniques of decentralisation to different degrees, this served mainly to reshape the role of centralisation and intermediaries, rather than to eradicate them. For example, uPort's *registry* stores a secure mapping between uPortID and their attributes and also relies upon central authorities as trust providers for identity attributes; the ShoCard central server is an intermediary that stores encrypted identity attributes and mediates between end-users and relying parties. Sovrin on the other hand embraces an open ecosystem of intermediaries (e.g., *agencies*, *trust anchors*).

So, while DLT applications often target the removal of the "middle man", this may not be a realistic goal in IdM applications due the context of identity maintaining a profound need for trust. Of course, this need for centralisation and intermediaries is not necessarily a bad thing: there are numerous examples of centralisation and intermediaries serving essential functions in an industry (see e.g., the SWIFT network). Elements of needed centralisation or intermediation in a decentralised IdM may comprise:

- Capturing additional authentication factors from end-users;
- Backing-up and recovering cryptographic keys;
- Providing a secure namespace to facilitate lookup of entities and services;
- Storing securely the information hash pre-images needed to validate digital signatures;
- Recovering compromised DLT-based identities.

The case of "The DAO" stands as an example of the risks of pursuing too much decentralisation in a system design. The DAO was designed as an Ethereum smart contract-based autonomous venture capital company but a flaw in its underlying code enabled an attacker to steal $50 million of the funding that it collected [12]. The research challenge for DLT applications in IdM is therefore to explore the balance between centralisation and decentralisation to create interoperable and privacy-respecting IdM that mitigates the risk of placing too much trust in any single authority.

### 6.2 Envisioned eco-systems of shareable identity attributes – but ad hoc trust

Support for the creation and sharing of identity attributes certified by 3[rd] parties is a design feature of each scheme we evaluated. In ShoCard, 3[rd] parties can certify attributes of an identifier; uPort and Sovrin support both self-attestation of attributes and those assigned by other entities.

Designing for reusable identity attributes aims to improve the granularity at which users can disclose identity information and promotes reuse of attributes. However, due to the lack of a central authority, trust of these attributes currently relies upon ad-hoc trust establishment between organisations. ShoCard and Sovrin propose a "web of trust" as the means by which attributes can be trusted. However, the challenges to design a web-of-trust are widely known where the network size is unbounded: difficulty to quantify trust beyond a first-degree relationship especially if any entity can vouch for any other, poor density of trust anchors on the network, lost or expired private keys, slow propagation of endorsement between users, etc. DLT does not address any of those challenges, but future research could focus on methods to achieve the building of trust and reputation between entities in the context of DLT identity attributes. This could be one way that DLT-based IdM responds to NSTIC [5] and delivers new interoperability in IdM.

### 6.3 If it isn't usable it isn't secure

Dhamija's explains in her "7 flaws of identity management" [13] that for users "identity management is not a primary goal". This has been reflected in the shrug that users have largely given to single-sign on solutions: the main user-facing proposition of IdM. This suggests that future IdM schemes with a novel technological underpinning but developed to the same blueprint of end-user interaction are unlikely to create widespread uptake. A principal tenet of human-computer interaction is to design systems based upon concrete knowledge of problems faced by end-users. So far we have seen that the schemes we evaluated are generally not accompanied by a novel vision of user interaction, and furthermore leave the perennial challenge to provide usable end-user key management [14] as largely unaddressed. Recent research has suggested that key management remains to be a principal source of concern for users of Bitcoin [15]. While the promising concept of key recovery was proposed in uPort and Sovrin, approaches to digital identity that remove central authorities and depend upon effective key management strategies from its users create the risk that non-technical users will be alienated by the technology; and when things go wrong those users will be unable to recover resources or reputation attached to lost keys.

## 7 Concluding remarks

Distributed Ledger Technology (DLT) is not a silver bullet solution for Identity Management (IdM). Our application of Cameron's evaluative framework provides an early glimpse of the current strengths and limitations of applying DLT to IdM. Future work in this nascent research area faces two particular hurdles:

Firstly, there is a noticeable lack of contextual understanding relating to the user experience elements of the schemes we encountered. Usability is a particularly pressing unknown since there appears to be a widespread assumption that users are equipped to conduct effective cryptographic key management, and would intuitively understand the implications of referencing identity data or attributes in a DLT.

Secondly, a tightening regulatory landscape for storing personal data. For example, the General Data Protection Regulation (GDPR) grants end-users new powers over personal data, and places new obligations upon data controllers. This creates a challenge for the design of immutable public ledgers that reference personal data, and that provide inherent transparency to data that they store.

Delaying the advance of new approaches to secure and trusted identities on the Internet was said to be an unacceptable course of action by the United States' NSTIC strategy [5]. This might be due to the concern that the online adage that "on the blockchain, nobody knows you're a fridge", may soon replace the prescience of Steiner's original cartoon.

**Table 1: A summary of uPort, ShoCard, Sovrin, and their relation to Cameron's laws of identity. Facebook Connect is provided for comparison. An unshaded table cell indicates that we found evidence that a scheme complied with a specific law, and a shaded cell indicates that we currently see no evidence that a scheme complies with a specific law.**

| Law | uPort | ShoCard | Sovrin | Facebook Connect |
| --- | --- | --- | --- | --- |
| 1 – User control and consent | User controls creation and disclosure of uPortIDs and can prove ownership of uPortID without a central authority. But attributes stored in *registry* may leak information. | User controls creation and disclosure of ShoCardIDs. Attributes are only accessible to a relying party by invitation of ShoCardID owner. But, ShoCard servers are necessary part of attribute validation protocol. | By design, users can choose which IDs are used and which attributes are revealed. A web of trust that could be reinforced by a reputation system helps protect users against deception. | Today, when using Facebook to log on to a service the user can choose which data will be shared by Facebook with the relying party. |
| 2 – Minimal disclosure for a constrained use | Users do not need to disclose personal data in order to create uPort identifiers for low value accounts. | ShoCardIDs are bootstrapped with a trusted identity document (e.g., a government ID). | Support of anonymous credentials based on zero-knowledge proofs allows users to share the information "least likely to identity [them] across multiple contexts" [1]. | A user can create an empty Facebook profile and progressively add identity information as is needed. |
| 3 – Justifiable parties | The JSON structure of attributes in the *registry* is visible to all, which may leak information to an honest-but curious attacker – even if encrypted. | ShoCardID only revealed to a relying party at the invitation of the ShoCardID owner. ShoCard servers may learn identity of relying parties. | Attributes are only accessible to relying parties that the user chooses, and to the agencies entrusted to act on their behalf. | Facebook always has access to the data stored on a user's Facebook profile whether the data is public or private. Facebook also creates and processes its own attributes e.g. relationships with friends. |
| 4 – Directed identity | Supports unidirectional sharing of identifiers between parties, but does not prevent entities broadcasting identifiers out of band e.g. on websites. | Supports unidirectional sharing of identifiers between parties, but does not prevent entities broadcasting identifiers out of band. | Omnidirectional identifiers are supported. | Omnidirectional identifiers are supported. A user's Facebook profile can be made public or private and profiles can be searched. |
| 5 – Design for a pluralism of operators and technology | Agnostic to the types of attributes that 3$^{rd}$ party identity providers create, yet use of a specific data format is encouraged in the *registry*. | Supports parsing of existing trusted credentials but relying parties must create bespoke integrations with ShoCard centralised servers for attribute validation. | Expects to build a market for intermediaries (*agencies*) between users and the Sovrin network. Some could be interfaces with other identity systems. | Only one identity provider: Facebook. Uses a bespoke method for authorisation to applications. But, Facebook has nearly 2 billion users. |
| 6 – Human integration | Provides a mobile application. Social cryptographic key recovery function shows promise. Unclear usability and user understanding of uPort privacy implications. | Provides a mobile application. The digital ID card metaphor is easy to understand. Unclear usability and user understanding of ShoCard privacy implications. | Implementation has been targeted so far towards the underlying technology, not the user experience. Unclear usability and user understanding of privacy. | Facebook is well known to users and usable interface to single-sign is provided. However, users may be unaware of privacy implications of Facebook Connect. |
| 7 – Consistent experience across contexts | User interaction driven by the mobile application. Consistently follows a QR code scanning paradigm for all uses. | User interaction driven by the mobile application. Consistently follows a QR code scanning paradigm for all uses. | Not clear. This will highly depend on the market of implementations of mobile device clients for the Sovrin network. | Consistent and experience via the 'login with Facebook' button. |

# Authors

**Paul Dunphy** is the lead researcher on the distributed ledger technology research theme at VASCO Data Security, based at their Innovation Centre in Cambridge, UK. Prior to joining VASCO he spent time at Atom Bank: the UK's first bank to deliver services entirely via mobile applications that pioneered the use of mobile biometrics. He joined during Atom's start-up phase and shaped the successful first launch of its mobile applications, which in total have processed close to £1 billion in customer deposits. He completed a Microsoft Research funded PhD at Newcastle University (UK) where his thesis focused on usable, secure and deployable user authentication. He has also spent time leading research projects at Microsoft Research and Nokia Research. His research interests are broadly at the intersection of privacy and security with human-computer interaction.

**Fabien Petitcolas** is research manager at VASCO Data Security, where he is contributing to grow and lead the recently established VASCO Innovation Centre. Prior to joining VASCO, Fabien spent fifteen years at Microsoft where he took various roles. He first became a member of the Security Group at Microsoft Research where he focused on digital watermarking. He later became head of Microsoft Research's intellectual capital development programmes, before becoming director for innovation at Microsoft Europe, supporting the company's presence in E.U. policy and political dialogue for and around innovation and R&D. Fabien received a PhD in computer science from the University of Cambridge (U.K.) under the guidance of Prof. Ross Anderson FRS FREng. His research interests include information hiding, an area where he has authored several publications and books, and, more recently, security issues related to identity management and user authentication.